\documentstyle[12pt,aasms4]{article}

\newcommand {\hii}{\ion{H}{2}}

\newcommand {\ha}{H$\alpha$}
\newcommand {\oiii}{[\ion{O}{3}]}
\newcommand {\sii}{[\ion{S}{2}]}
\newcommand {\degr}{\arcdeg}
\newcommand {\amin}{\arcmin}
\newcommand {\asec}{\arcsec}
\newcommand {\rahr}{$^h$}
\newcommand {\ramin}{$^m$}
\newcommand {\rasec}{$^s$}
\def \ampt      {\farcm}

\newcommand {\et}{{et al.}}
\newcommand {\eg}{{e.g.,}}

\newcommand {\kms}{km~s$^{-1}$}

\newcommand {\rosat}{{\it ROSAT\ }}

\newcommand {\chandra}{{\it Chandra\ }}
\newcommand {\asca}{{\it ASCA\ }}
\begin{document}

\title {Resolving SNR 0540$-$6944 from LMC X-1 with Chandra}
\author{Rosa Murphy Williams\altaffilmark{1}}
\author{Robert Petre\altaffilmark{2}}
\author{You-Hua Chu\altaffilmark{3}}
\and
\author{C.-H. Rosie Chen\altaffilmark{4}}

\altaffiltext{1}{National Resarch Council Associate, 
	NASA's GSFC, code 662, Greenbelt, MD 20771, USA;
	rosanina@lhea1.gsfc.nasa.gov}
\altaffiltext{2}{NASA's GSFC, code 662, Greenbelt, MD 20771, USA;
	rob@hatrack.gsfc.nasa.gov}
\altaffiltext{3}{Astronomy Department, University of Illinois, 
	1002 W. Green Street, Urbana, IL 61801, USA;
	chu@astro.uiuc.edu}
\altaffiltext{4}{Astronomy Department, University of Illinois, 
           	1002 W. Green Street, Urbana, IL 61801, USA;
	c-chen@astro.uiuc.edu}

\begin{abstract}

We examine the supernova remnant (SNR)  0540$-$697 in the Large Magellanic
Cloud (LMC) using data from the \chandra ACIS.  The X-ray emission from
this SNR had previously been hidden in the bright emission of nearby X-ray
binary LMC X-1;  however, new observations with \chandra can finally
reveal the SNR's structure and spectrum. We find the SNR to be a
thick-shelled structure about 19 pc in diameter, with a brightened
northeast region.  Spectral results suggest a temperature of 0.31 keV and
an X-ray luminosity (0.3-3.0 keV) of 8.4 $\times 10^{33}$ erg s$^{-1}$.  
We estimate an age of 12,000-20,000 yr for this SNR, but note that this
estimate does not take into account the possibility of cavity expansion or
other environmental effects.

\end{abstract}


\section{Introduction}

Recent observations of the Large Magellanic Cloud (LMC) have opened up a
new era in near-extragalactic studies of supernova remnants (SNRs).  
Thanks to advances in instrumentation over the last decade, we have been
able to observe this large and varied sample of remnants without the
difficulties in distance determination and obscuration for Galactic
remnants, or those in resolution and sensitivity for ones in more distant
galaxies. This has allowed us both to examine individual remnants in
detail (\eg\ Williams \et\ 1997, 1999a; Chu \et\ 1997, 2000) and as a
group (\eg\ Williams \et\ 1999b; Williams 1999) to explore their
interactions with the surrounding interstellar medium (ISM).

The completeness of the LMC sample, however, is a vexing question. 
Many SNRs lack one or more of the traditional SNR signatures (high 
\sii/\ha\ ratios, nonthermal radio emission, and X-ray emission) and 
must be identified through other means (\eg\ Chu 1997).  
New LMC SNRs continue to be found (\eg\ Smith \et\ 1994; Chu \et\ 1993, 
1995, 1999).  Each  discovery not only adds to the list of known SNRs, but 
provides additional insight into how to detect other such SNRs. This has 
profound implications for both the completeness of the LMC sample and 
the assemblage of samples from other galaxies, and thus for our 
understanding of the contribution from SNRs to the energetics and
dynamics of the ISM.

One of these discoveries, SNR 0540$-$6944, may prove particularly
informative in this regard. The SNR, discovered by Chu \et\ (1997), is
difficult to observe in optical and radio due to emission from the
surrounding \hii\ region N159; the X-ray emission was similarly obscured
by that from the nearby X-ray binary LMC X-1.  The SNR's expansion was
serendipitously uncovered in optical echelle observations, and its SNR
nature confirmed by using soft-band (0.1-1.0 keV) \rosat PSPC images to
separate the SNR's thermal emission from the much harder emission of LMC
X-1. However, the SNR remained spatially and spectrally confused with LMC
X-1 at the resolution of the \rosat\ observations.

The unprecedented combination of spatial and spectral resolution and
sensitivity from the Chandra (formerly AXAF) X-ray Observatory and its
instruments allows us, for the first time, to get a picture of SNR
0540$-$6944 distinct from that of LMC X-1.  Using these data, we are able
to present images and spectra from the object itself.  The results 
confirm the SNR identification of Chu \et\ 1997; provide spatial and
spectral data on the SNR itself; and illustrate the information that
may be gleaned by separating objects whose close proximity confuses their
emission. It provides a striking demonstration of the many new areas of
investigation made possible by \chandra's power, even in a brief
observation.

\section{Observations}

LMC X-1 was observed as a calibration source by \chandra in late 1999 and
early 2000.  For our spectral and soft-band imaging studies, we used the
Advanced CCD Imaging Spectrometer (ACIS) on-axis observation (sequence
number 490002, 6.5 ksec) to avoid additional problems with off-axis
distortions.  Additionally, the target was focused on chip S3, a
back-illuminated chip, which allowed for slightly greater spectral
sensitivity, while avoiding the problems with radiation damage
to which the front-illuminated chips were subject (Orbital Calibration
reports, ASC, 1999).  Only about 3 ksec of data are available for
analysis; other datasets available for this object were off-axis,
contained processing errors, or both.  The ACIS (using a back-illuminated
chip) has an angular resolution of 1\asec, an energy resolution of 100 eV
at 1.0 keV (E/$\Delta$E =9), an energy range of 0.2-10 keV, and an
effective area of 600 cm$^{-2}$ at 1.0 keV (\chandra Observatory Guide,
ASC, 1997).  Data reduction and analysis were performed using the CIAO 
(Chandra X-ray Center software) and FTOOLS, XSPEC, and XIMAGE
data-processing routines (Arnaud 1996).

\section{Analysis}

While SNR 0540$-$6944 can be spatially separated from LMC X-1 using the
ACIS-S, the flux from the SNR is very low, and close to the background
level.  As a result, both spatial and spectral analysis are difficult. In
addition, due to frame transfer effects, a ``spike" of emission from LMC
X-1 intersects the circle of emission from the SNR.  These factors
complicate attempts to isolate the the SNR itself from its much brighter
neighbor.

\subsection{Morphological Analysis}

A first look at the image of this region on the \chandra ACIS-S is
disappointing. Even using the S3 back-illuminated chip, more sensitive to
low-energy emission, the SNR's presence is barely detectable next to that
of LMC X-1.  However, the emission from SNR 0540$-$6944 is likely to fall
largely in the energy band between 0.1 and 3 keV, as is expected for a
thermal plasma.  We therefore make exposure-corrected images in the soft
(0.3-3.0 keV) and hard (3.0-9.0 keV) bands and compare them.  (Fig. 1a-b;
note that we have used a low-energy cutoff of 0.3 keV to reduce
contributions from the soft X-ray background.  No counts are expected from
the LMC below this cutoff due to the intervening column density.) We do
indeed see emission at the position of SNR 0540$-$6944 in the soft image;
but it is still overwhelmed by that from LMC X-1.

In order to eliminate some of the contamination by LMC X-1, we subtract
the hard map from the soft map, thus removing emission from areas where
hard X-rays dominate.  In order to bring up the contrast in the resulting
map, we divide by a total map.  What remains is a ``softness ratio" map in
the form (soft - hard) / (soft + hard).  This can be used to discern the
structure of the soft emission, presumably that from the SNR (Fig. 1c-d).
What is thus revealed is a roughly circular structure centered at J2000.0
coordinates 05\rahr 40\ramin 05\rasec, -69\degr 44\amin 07\asec.  The
circle has a rough diameter of $\sim$1$^{\prime}.25$, or 19 pc at the
distance to the LMC (50 kpc). The emission is distributed over the
remnant, suggesting a thick-shelled structure.  The remnant is
considerably brighter in a small ($\sim$9\asec\ radius) region to the
northeast.

For the purpose of optical comparison we obtained archival {\it Hubble
Space Telescope}\ images in the H$\alpha$ and \oiii\ emission lines (Fig.
1e-f; PEP ID 6535; H$\alpha$: four 300 sec exposures; \oiii: four 230 sec
exposures).  These images show a roughly circular, highly filamentary
structure amidst the emission from the rest of the N159 region. This
structure corresponds very well to the position and extent of X-ray
emission revealed by the ACIS ratio image, suggesting that the optical
emission comes from the cooling shell of the SNR.

\subsection{Spectral Analysis}

In order to separate the SNR emission from that of LMC X-1, it becomes
useful to consider the emission of the X-ray binary itself.  In this we
are aided by the availability of data from the \asca\ SIS (ad43004000,
12.5 ksec).  \asca\ is insensitive at energies below 0.7 keV, so the
contribution from the SNR to the data is expected to be minimal.  We can,
therefore, take the \asca\ data as representative of LMC X-1 alone.

We therefore approached the problem using four separate spectra. Two were
from \asca\ observations using the SIS0 and SIS1 instruments; the region
covered includes both LMC X-1 and SNR 0540$-$6944.  A third spectrum was
extracted from \chandra\ ACIS-S data, similarly covering a region
including both LMC X-1 and SNR 0540$-$6944.  A fourth spectrum, also
extracted from \chandra\ ACIS-S data, covered SNR 0540$-$6944 only.

From previous studies of LMC X-1 (\eg\ Schlegel \et\ 1994) we know that
the spectrum of this X-ray binary is well represented by the combination
of disk-blackbody and power-law models.  Previous studies (\eg\ Schmidtke,
Ponder, \& Cowley 1999) suggest that there are no significant long-term
variations in the spectrum of LMC X-1, allowing us to meaningfully compare 
datasets taken at different times.  We expect the SNR
contribution to be reasonably well modeled by a thermal plasma model
(Raymond \& Smith 1977).  Our combined model for the region, then, has
four components: one for photoelectric absorption (based on Morrision \&
McCammon 1983), applied to a combined Raymond-Smith, disk-blackbody and
power-law model. Abundances for the Raymond-Smith model were set to 0.3
solar, as appropriate to the ISM of the LMC (Russel \& Dopita 1992).

This model was simultaneously fit to our four spectra.  The model
parameters were linked, with the exceptions of the normalizations, which
were allowed to fit independently to the four spectra.  For the \asca\
spectra, the Raymond-Smith normalizations were set to zero, as little
thermal contribution was expected. Likewise, the normalizations for the
disk-blackbody and power-law components were set to zero for the \chandra\
spectrum of 0540$-$6944 alone, as the contributions from LMC X-1 were
expected to be minimal.

The best-fit parameters are given below, with the 90\% confidence ranges
given in parentheses.  The best fit for the X-ray absorption of the
region, $N_H=7.2 (6.9-7.5) \times 10^{21}$ cm$^{-2}$, is similar to that
found by Schlegel \et\ (1994) and Schmidtke \et\ (1999) for LMC X-1; it is
also somewhat atypically high for the LMC.  The best-fit parameters for
LMC X-1 are a blackbody temperature of kT$_{bb}=$0.82 (0.81-0.83) and a
power-law index of $\Gamma$=2.3 (2.2-2.4), again consistent with the
results from Schlegel \et\ (1994) and Schmidtke \et\ (1999).  The thermal
plasma component yields a best-fit temperature of kT$_{rs}$=0.31
(0.18-0.43), reasonable for an older SNR (Fig. 2). Based on these spectral
results, we computed a flux from the SNR in the 0.3-3 keV range of 2.8
$\times 10^{-14}$ erg cm$^{-2}$ s$^{-1}$, and a luminosity in the same
range of 8.4 $\times 10^{33}$ erg s$^{-1}$.  When we look at the data for
SNR 0540$-$6944 and LMC X-1 combined, we find a flux in the 0.3-3 keV
range of 1.9 $\times 10^{-12}$ erg cm$^{-2}$ s$^{-1}$ and a luminosity of
5.7 $\times 10^{35}$ erg s$^{-1}$. Thus, within the energy range
specified, the SNR contributes about 1.5\% of the X-ray flux from this
region.

\section{Discussion}

Morphologically, the SNR is without a sharply defined shell in X-rays.  
The large and distributed X-ray structure, interior to much of the optical
emission, implies an older, Sedov-stage SNR.  This picture is strengthened
by the relatively low temperature and low luminosity of the X-ray
emission.  The velocity of $\sim$180 \kms\ found by Chu et al. (1997)
indicates a slow expansion consistent with this picture.  If this velocity
is representative of the actual expansion velocity of the remnant, and the
SNR is undergoing Sedov-like expansion, we would expect the shock velocity
to be $v_{shock} = 4/3 v_{exp} = 240$ km s$^{-1}$.

This shock velocity, in turn, would imply a temperature of $kT = 3/16 \mu
v_{shock}^2$, where $\mu$ is the mean molecular weight (assumed 1.1$m_H$).
For the shock velocity given above, this gives approximately 0.12 keV.  
The temperature derived from X-ray spectral fits is somewhat higher; the
X-ray temperature would imply $v_{exp} = 280$ (220-340) km s$^{-1}$.  
Similar discrepancies have been noted for other LMC remnants (Williams 
1999). One possible explanation for this discrepancy is that the highest-velocity
material may be too faint to show up clearly in the echelle spectrum.  
Another possibility is that the expansion velocity is indeed accurately
reflected by the echelle spectroscopy and that this slow shock speed is
insufficient to produce X-ray emission at the shock front. In this latter
case, the remnant may have entered the shell-forming phase, as evidenced
by the pronounced H$\alpha$ shell.  The observed X-rays in such a case are
likely to be ``fossil"  radiation produced by the cooling of gas shocked
to high temperatures earlier in the remnant's expansion.

Using the Sedov equation and the velocity from Chu et al. (1997), and
assuming constant external density, we find an age of $\sim$20,000 yr for
this SNR.  The velocity derived from the X-ray temperature would give an
age of $\sim$12,000 yr.  These must be regarded as only approximate
figures.  For instance, if the SNR is expanding in the wind-blown bubble
formed by its progenitor - a quite plausible scenario, as the remnant is
within an \hii\ region - it may very well be considerably younger than
these estimates. Given this estimate of age, we do not expect a causal
connection between SNR 0540$-$6944 and LMC X-1.  The projected distance
between the center of the SNR and LMC X-1 is about 1\ampt75, or a minimum
of 26 pc separation.  To reach this distance within the estimated
age of the SNR, the compact object would have had to travel at a constant
speed of over 1200 km s$^{-1}$.

Our fits to the X-ray spectrum use a normalization constant directly
related to the emissivity, $K = 10^{-14} \int n_e n_H dV / (4 \pi D^2)$.
Here D is the distance to the remnant, $n_e$ and $n_H$ the electron and
particle densities (we assume $n_e = 1.1 n_H$), and $V$ the remnant
volume.  This allows us to make estimates of the density, energy and
pressure within the X-ray emitting gas of this remnant.  Using the fitted
value of $K = 5.66 \times 10^{-4}$ and assuming a volume filling factor
for the gas of 10\% (corresponding to a shell thickness of about 0.33 pc,
3\% of the radius, a reasonable value for a middle-aged remnant), we 
calculated the gas density at about $n$=1.2
cm$^{-3}$.  Using the formula $E_{th} = (3/2) N kT$, where N is the total
particle number in the hot cavity, we find a thermal energy of about
$10^{49}$ erg, suitable for a remnant in which much of the remaining
energy is tied up in the kinetic energy of expansion.  The thermal
pressure of the hot gas, according to $P=nkT$, is about 6 $\times
10^{-10}$ dyne cm$^{-2}$.  These figures should, of course, only be
regarded as rough estimates, as there is perhaps an order of magnitude
uncertainty in the actual volume occupied by the hot gas, as well as
additional uncertainties in the fitted temperature and emissivity.

The reason for the brightening of X-rays in the northeast section of the
remnant is unclear.  A search for timed emission revealed no significant 
peaks in the power spectrum; however, given the short exposure
and high background for this observation, further investigation is
indicated.  The X-ray brightening occurs near an optical ``knot" of bright
emission, and may indicate a region where the SNR shock is
encountering denser material.

In summary, we find that Chandra data are sufficient to distinguish SNR
0540$-$6944 from the emission of the nearby X-ray binary LMC X-1.  While
these findings are preliminary, as there remain uncertainties in the
spatial and spectral responses, they are still informative.  Given that
this SNR has a lower luminosity than most LMC SNRs (eg Williams 1999), the
results point out the capacity of Chandra to distinguish faint objects in
confused regions.  As observations continue, we may uncover an entire
population of low-luminosity, older SNRs, in the LMC and even in more
distant galaxies.  This has the potential to substantially increase our
estimates of SNR rates, which to date have been largely based on the
fraction of SNRs that are more readily detected.  SNR 0540$-$6944 also
suggests the possibility of finding previously undetected SNRs within the
crowded environs of \hii\ regions, where we would indeed expect a high
population of Type II SNRs.

\acknowledgements 
The authors thank Eric Schlegel for his helpful comments
as referee, as well as the CXC team for their work in making the data and
analysis tools available. We acknowledge the support of the National Resarch 
Council (RMW) and NASA ADP Grant NAG 5-7003 (RMW \& YHC).

\clearpage

\begin{figure}

\caption{Image of SNR in (a) ACIS soft map; (b) ACIS hard map;
(c) ACIS ratio map; (d) ACIS ratio map scaled to HST field, with
contours at 2, 4, 6, 8, 10, 12$\sigma$ over background; 
(e) HST H$\alpha$ with ACIS 2, 6, 10$\sigma$ contours; 
and (f) HST \oiii\ with ACIS 2, 6, 10$\sigma$ contours}

\caption{Spectral fits to X-ray data from (top two) \asca SIS for
LMC X-1; (middle) \chandra ACIS for LMC X-1 and SNR; (bottom)
\chandra ACIS for SNR}

\end{figure}


\clearpage

\begin{references} 
\reference{A96} Arnaud, K. A. 1996, Astronomical Data Analysis Software
     and systems V, eds Jacoby, G. \& Barnes, J., ASP Conf. Series 101, p. 17
\reference{C+00} Chu, Y.-H., Kim, S., Points, S. D., Petre, R., \&
     Snowden, S. L. 2000, AJ, in press (May issue)
\reference{C+97} Chu, Y.-H., Kennicutt, R. C., Snowden, S. L., Smith,
     R. C., Willams, R. M., \& Bomans, D. J. 1997, PASP, 109, 554
\reference{C97} Chu, Y.-H. 1997, AJ, 113, 1815
\reference{C+95} Chu, Y.-H., Dickel, J.R., Staveley-Smith, L., Osterberg, J., 
       Smith, R.C. 1995, AJ, 109, 1729
\reference{C+93} Chu, Y.-H., Mac Low, M.-M., Garcia-Guillermo, G., Wakker, B., 
     \& Kennicutt, R. C.  1993, ApJ, 414, 213
\reference{MM83} Morrison, R., \& McCammon, D. 1983, ApJ, 270, 119
\reference{RS77} Raymond, J. C., \& Smith, B. W. 1977, ApJS, 35, 419 
\reference{S+94} Schlegel,  E. M., Marshall, F. E., Mushotzky, R. F., Smale, A. P., 
     Weaver, K. A., Serlemitsos, P. J.,  Petre, R., \& Jahoda, K. M. 1994, ApJ, 422, 243
\reference{S+99} Schmidtke, P. C., Ponder, A. L., \& Cowley, A. P. 1999, AJ, 117, 1292
\reference{S+94} Smith, R.C., Chu, Y.-H., Mac Low, M.-M., Oey,
     M. S., Klein, U. 1994, AJ, 108, 1266 
\reference{W+97} Williams, R. M., Chu, Y.-H., Dickel, J. R., Beyer, R., 
    Smith, R. C., Petre, R., \& Milne, D. K. 1997,  ApJ, 480, 618 
\reference{W+99a} Williams, R. M., Chu, Y.-H., Dickel, J. R., \& Smith, R. C.
    1999a, ApJ, 514, 798  
\reference{W+99b} Williams, R. M.,  Chu, Y.-H., Dickel, J. R., Petre, R.,
   Smith, R. C., \& Tavarez, M. 1999b, ApJS, 123, 467 
\reference{W99} Williams, R. M.,  1999, Ph.D. Thesis, University of Illinois at Urbana
\reference{}
\end{references}
\end{document}